# Macroelement modeling of shallow foundations

Chatzigogos, C., T., Pecker, A., Salençon, J.

*Laboratoire de Mécanique des Solides CNRS UMR 7649, Département de Mécanique, École Polytechnique, Palaiseau 91128 Cedex FRANCE*

*ABSTRACT. The present article is concerned with the development of a macroelement model for shallow foundations. The model aims at serving as a practical tool for quick and precise non-linear dynamic analyses of structures, taking into account the soil-structure interaction non-linearities that take place at the foundation level. After a brief overview of some existing macroelement models we outline the principles followed in the development procedure. The macroelement is formulated with introduction of a non-linear constitutive law written in terms of a set of generalized force and displacement parameters. The linear part of this constitutive law is reproduced by the dynamic impedances of the foundation. The non-linear part comprises two mechanisms. The first one, of material origin, is due to the irreversible elastoplastic behavior of the soil. It is described with a bounding surface hypoplastic model, particularly adapted for the description of cyclic soil response. An original feature of the formulation with respect to previous macroelement models is that the bounding surface of the plasticity model is considered independently of the surface of ultimate loads of the system. The second mechanism, of geometric origin, is due to the conditions of unilateral contact at the soil-footing interface, allowing for uplift of the footing. This mechanism is perfectly reversible and non-dissipative. It can thus be described by a phenomenological non-linear elastic model. The macroelement is qualitatively validated by application to soil-structure interaction analyses of simple real structures.*

**Keywords:** shallow foundations, displacement-based design, non-linear dynamic analysis, plasticity, uplift.

# 1 Introduction

## 1.1 Definition of the "macroelement"

We are presenting in this article a new formulation for the modeling of shallow foundations of structures using the concept of macroelement. The macroelement can be viewed as a practical tool, which allows for efficient dynamic analyses of structures with consideration of the non-linear soil-structure interaction effects arising at the foundation level. As a generic example, we consider the configuration of Figure 1, where a soil-foundation-superstructure system is subject to a dynamic excitation at the soil bedrock denoted by $\underline{\ddot{u}}$. The problem viewed in its entirety entails a number of non-linearities such as the irreversible elastoplastic soil behavior and the unilateral soil-structure interface conditions leading to uplift of the structure, both of which render its numerical treatment within a classical finite-element framework, delicate and particularly expensive. Furthermore, the dynamic nature of the problem makes it even more challenging: the model needs to be able to accommodate for an accurate description of the wave propagation and radiation phenomena (the second arise as waves emanate from the foundation towards the infinite extremities of the soil medium) and it is clear that fully non-linear dynamic analyses in the time domain for three-dimensional configurations remain beyond the reach of conventional computational capacities. In such a setting, the concept of "macroelement" is introduced by replacing the whole foundation-soil system by a singe element which is placed at the base of the superstructure and aims at reproducing the non-linear soil-structure interaction effects taking place at the foundation level, on the overall response of the superstructure. Consequently, this element exhibits a non-linear "constitutive law", which links some generalized force parameters with the corresponding cinematic ones. The generalized force and displacement parameters are chosen in such a way so as to be coherent with those adopted for the superstructure model.

## 1.2 Existing macroelement models for shallow foundations

The concept of "macroelement" has been initially introduced in the context of shallow foundations by Nova & Montrasio [1]. Based on a number of experimental tests performed on a perfectly rigid strip footing resting on a frictional soil and subject to an eccentric and inclined force, Nova & Montrasio calibrated a global elastoplastic model with isotropic hardening for the entire soil-foundation system. The model was written in terms of the resultant vertical and horizontal force and moment acting on the footing normalized by the maximum supported vertical force and was used for the prediction of the footing displacements for quasistatic monotonic loading. The rugby-ball shaped ***surface of ultimate loads*** of the system was identified as the ***yield surface*** of the plasticity model. This surface is schematically presented in Figure 2.

The model of Nova & Montrasio was modified in the first place by Paolucci [2] for application to structures subjected to real dynamic loading and further extended by Pedretti [3] for a more accurate description of the system behavior under cyclic loading. Crémer [4] and Crémer *et al.* [5], [6] presented an advanced macroelement model where two separate mechanisms in coupling have been introduced: the first one referring to the material non-linearity of the system due to the irreversible elastoplastic soil behavior and the second one describing the uplift of the footing due to the unilateral contact conditions on the interface. For the description of the system behavior under cyclic loading Crémer formulated a plasticity model with isotropic and cinematic hardening. Le Pape & Sieffert [7], [8] presented a macroelement model similar to the one proposed by Nova & Montrasio particularly oriented for earthquake engineering applications and based on thermodynamical principles. Several macroelement models have also been proposed in the context of the off-shore industry for a variety of soil conditions and foundation geometries. Efforts have been performed to obtain global models of shallow foundations by considering decoupled Winkler springs attached at the foundation interface as in [9], [10] that are characterized by an elastoplastic contact-breaking law. The advantage of such formulations is that they allow obtaining the global system response by integration of the local spring response, which can be achieved analytically. On the other hand, they are subject to all type of constraints associated with the Winkler decoupling hypothesis, such as the difficulty to calibrate model parameters. We note finally that the concept of macroelement has been applied to other types of geotechnical problems as, for instance, the dynamic response of gravity walls (*cf.* [11]). The non-exhaustive Table 1 presents an overview of some of the existing macroelement models for shallow foundations.

## 2 Model formulation

## 2.1 Definition of generalized forces and displacements

The modeling procedure is initiated with the definition of the generalized forces and displacements, in terms of which the "constitutive" equations for the macroelement are written. In the following, we will consider the particular case of a perfectly rigid circular footing resting on the surface of the soil. The perfect rigidity of the footing allows knowledge of the movement of all its points if the movement of a single point is known. We thus consider the resultant forces in the vertical and horizontal direction and the resultant moment acting at the center of the footing. We also consider the corresponding cinematic parameters: vertical and horizontal displacement and angle of rotation at the center of the footing. A planar loading will be considered in the following, although the model presented is adapted to allow for easy extension to fully three-dimensional configurations. The examined footing is presented in Figure 3.

The "constitutive" equations of the macroelement will be written in terms of the force and displacement parameters presented in Figure 3, which are introduced normalized according to the following scheme:

(1)
$$\underline{Q} = \begin{bmatrix} Q_N \\ Q_V \\ Q_M \end{bmatrix} = \begin{bmatrix} N/N_{\max} \\ V/N_{\max} \\ M/DN_{\max} \end{bmatrix}, \quad \underline{q} = \begin{bmatrix} q_N \\ q_V \\ q_M \end{bmatrix} = \begin{bmatrix} u_z/D \\ u_x/D \\ \theta_y \end{bmatrix}$$

In (1), $D$ is the diameter of the circular footing and $N_{\max}$ is the maximum vertical force supported by the footing. The introduced normalization leads in the following expression for the work of the force parameters:

(2)
$$\mathcal{W}(\underline{Q},\underline{q}) = \underline{Q} \cdot \underline{q} = Q_i q_i =$$
$$\frac{1}{N_{\max}D}(Nu_z + V_x u_x + M_y \theta_y) = \frac{\mathbf{W}}{N_{\max}D}$$

In other words, the total work $\mathbf{W}$ attributed in the system is normalized by the fixed quantity $DN_{\max}$. We also note that if the force and displacement increments are related by introduction of a general stiffness matrix as in the following expression:

(3)
$$\begin{bmatrix} \dot{Q}_N \\ \dot{Q}_V \\ \dot{Q}_M \end{bmatrix} = \begin{bmatrix} K_{NN} & K_{NV} & K_{NM} \\ K_{VN} & K_{VV} & K_{VM} \\ K_{MN} & K_{MV} & K_{MM} \end{bmatrix} \begin{bmatrix} \dot{q}_N \\ \dot{q}_V \\ \dot{q}_M \end{bmatrix},$$

then, following (1), the elements of the stiffness matrix are subject to the following normalization scheme:

(4)
$$\underline{\underline{\mathcal{K}}} = \begin{bmatrix} (D/N_{\max})\mathbf{K}_{NN} & (D/N_{\max})\mathbf{K}_{NV} & (1/N_{\max})\mathbf{K}_{NM} \\ (D/N_{\max})\mathbf{K}_{VN} & (D/N_{\max})\mathbf{K}_{VV} & (1/N_{\max})\mathbf{K}_{VM} \\ (1/N_{\max})\mathbf{K}_{MN} & (1/N_{\max})\mathbf{K}_{MV} & (1/DN_{\max})\mathbf{K}_{MM} \end{bmatrix}$$

The quantities $K_{ij}$, $i,j = N,V,M$ in (3), represent the normalized elements of the stiffness matrix. Similarly, the quantities $\mathbf{K}_{ij}$, $i,j = N,V,M$ in (4) represent the dimensional elements of the stiffness matrix of the real system.

## 2.2 Structure of the macroelement model

The basic remark concerning the structure of the model is that the global behavior of the system reproduced by the macroelement is actually the result of the combination of the soil and the soil-footing interface properties. The macroelement should thus reflect the rigidity and strength characteristics of the soil as well as the strength characteristics of the soil-foundation interface. The different possibilities existing for these properties will give rise to different macroelement formulations. The macroelement presented herein is developed for applications in earthquake engineering, so we adopt the following assumptions:

*a.* The applied seismic loads being in general of very short duration (of the order of magnitude of few seconds), the soil response will correspond to **undrained conditions** of loading. It will thus give rise to a Tresca strength criterion, which may be incorporated within an **associated** plasticity model.

*b.* The soil-foundation interface is a no-tension interface that allows for uplift of the footing. This is an essential and desirable feature for applications in earthquake engineering, where it has been commonly observed that the uplift taking place at the foundation level acts as a seismic isolation mechanism for the superstructure. Interface strength criteria that satisfy this condition are the perfectly rough no-tension interface, the Tresca interface without resistance to tension, the Coulomb interface with zero cohesion *etc*. For the needs of the present developments we will retain the perfectly rough no-tension interface.

The soil and the interface strength criteria are combined in the plane directly below the footing and can be represented in a $\sigma - \tau$ diagram, $\sigma$ denoting the normal and $\tau$ the tangential component of the traction on the plane, as in Figure 4(a). We also note that the considered criteria for the soil and the soil-footing interface give rise to a surface of ultimate loads for the system which is represented in Figure 4(b) in the space of the generalized force parameters $(Q_N, Q_V, Q_M)$. (*cf.* [18]). We insist in the fact that this surface is obtained as a combined result of the soil and the soil-footing interface strength criteria.

Given the aforementioned assumptions for the system behavior, the passage to the macroelement is performed based on the following remarks:

*a.* The two mechanisms governing the system behavior, *i.e.* the mechanism of uplift and the mechanism of soil plasticity will be modeled independently and they will be incorporated in the macroelement. This will allow recuperating one mechanism if the other one is deactivated.

*b.* Figure 4(a) reveals that the nature of the two mechanisms is diametrically different. The soil plasticity mechanism concerns a dissipative process accounting for the irreversible elastoplastic soil behavior. It can be modeled within the macroelement by an associated plasticity model formulated in terms of the generalized forces and displacements. If we isolate this mechanism by considering that no uplift is allowed at the soil-footing interface, then the obtained global yield surface can be approximated by an ellipsoid in the space of the parameters $(Q_N, Q_V, Q_M)$ centered at the origin, as it has been shown in [19].

*c.* On the contrary, the uplift mechanism, pertaining to a non-linearity of geometric nature, concerns a non-dissipative reversible process. Since the introduction of the macroelement is performed by substituting the entire soil-foundation geometry by a single point, the possibility of modeling the uplift mechanism by taking into account the change of the system geometry is *a priori* excluded. The uplift mechanism can however be modeled within the macroelement with a phenomenological non-linear elastic model written in terms of the generalized forces and displacements, which respects its reversible and non-dissipative nature and reproduces in a phenomenological way the apparent reduction of the foundation stiffness or the apparent heave of the footing center as uplift is initiated *etc*.

*d.* The coupling of the two mechanisms within the macroelement should provide the surface of ultimate loads of the system, presented in Figure 4(b). But there in no reason whatsoever, to use the ultimate surface of the system for the definition of any of the two separate mechanisms. This is especially the case for the yield surface of the plasticity model, which has been traditionally identified in all previous macroelement models with the ultimate surface of the system. In the examined case, such an assumption can not be justified since the ultimate surface is obtained as the combined result of both mechanisms.

Following the aforementioned remarks, the structure obtained for the macroelement is presented in Figure 5(a) (plane $Q_N - Q_V$) and in Figure 5(b) (plane $Q_N - Q_M$). The plasticity model is introduced by an ellipsoidal yield surface with an associated flow rule and the uplift model by a cut-off at $Q_N = 0$. In the interior of the yield surface, the elastic response of the system remains linear before uplift is initiated and turns into non-linear after uplift initiation. The region $Q_N < 0$ corresponds to a situation where the footing is totally detached from the soil and its treatment will not be included in the model.

Concerning the influence of the horizontal force on the uplift model, we will assume, following Crémer (*cf.* [4]), that for soils of a certain depth, the horizontal force has no effect on the uplift response of the system. This leads to a formulation of the uplift model with respect to the parameters $Q_N, Q_M$ only.

The advantage of the proposed structure is that it induces the very simple rheological model presented in Figure 6, for which the decomposition of the displacement increment into an elastic and plastic part can be introduced:

(5) $$\underline{\dot{q}} = \underline{\dot{q}}^{\text{el}} + \underline{\dot{q}}^{\text{pl}}$$

In the following sections we describe in detail the plasticity and uplift models implemented in the macroelement.

## 2.3 Non-linear elastic model for uplift

We define initially the uplift model, which is a phenomenological non-linear elastic model. This allows incorporating the linear elastic part of the behaviour of the system into the uplift model. The model will be formulated independently of any plastic soil behaviour (we thus consider that the plasticity mechanism is deactivated). We initially introduce an incrementally linear relationship linking the increment of forces with the increment of the elastic displacements:

(6) $$\underline{\dot{Q}} = \underline{\underline{K}} \, \underline{\dot{q}}^{\text{el}}$$

In (6), $\underline{\underline{K}} = K_{ij}$, $i,j = N,V,M$ is the tangent elastic stiffness matrix of the system, with elements that are not constant in general. Our goal will be to formulate a purely phenomenological model describing the uplift of the footing, in such a way so that the elements $K_{ij}$, $i,j = N,V,M$ are functions of the generalized elastic displacements:

(7) $$K_{ij} = K_{ij}\left(\underline{q}^{\text{el}}\right)$$

### 2.3.1 Elastic stiffness matrix before uplift initiation

Before uplift initiation the system response is linear elastic and the matrix $\underline{\underline{\mathcal{K}}}$ assembles the static impedances of the foundation. For a circular footing with planar base that rests on the soil surface the coupling terms are negligible (*cf.* [20]), so the matrix $\underline{\underline{\mathcal{K}}}$ is written:

$$(8) \quad \underline{\underline{\mathcal{K}}} = \begin{bmatrix} \widetilde{K}_{NN} & 0 & 0 \\ 0 & \widetilde{K}_{VV} & 0 \\ 0 & 0 & \widetilde{K}_{MM} \end{bmatrix}$$

In (8), the quantities $\widetilde{K}_{NN}, \widetilde{K}_{VV}, \widetilde{K}_{MM}$ depend on the foundation geometry and on the elastic parameters of the soil.

### 2.3.2 Elastic stiffness matrix during uplift

The elastic stiffness matrix during uplift will be calibrated using finite element solutions of the uplift of a footing resting on a purely elastic soil. The model will be independent of $Q_V$ and we consider that uplift is initiated when the moment $|Q_M|$ applied on the footing exceeds (in absolute value) a certain value $|Q_{M,0}|$:

*Before uplift*: $\quad |Q_M| < |Q_{M,0}| \Rightarrow Q_M = \widetilde{K}_{MM}\, q_M^{el}$

*Uplift initiation*: $\quad |Q_M| = |Q_{M,0}| \Rightarrow q_{M,0}^{el} = \dfrac{Q_{M,0}}{\widetilde{K}_{MM}}$

We note that the quantity $|Q_{M,0}|$ is a function of the vertical force applied on the footing. For circular footings resting on elastic half-spaces, Wolf [21] proposed:

$$(9) \quad Q_{M,0} = \pm \frac{Q_N}{3} = \pm \frac{1}{3}\widetilde{K}_{NN}\, q_N^{el}$$

So the elastic angle of rotation at the instant of uplift initiation is written:

$$(10) \quad \left|q_{M,0}^{el}\right| = \frac{1}{\widetilde{K}_{MM}}\left(\frac{1}{3}\widetilde{K}_{NN}\, q_N^{el}\right)$$

For the calibration of the elastic tangent stiffness matrix during uplift and in absence of specific results for circular footings, we will use the numerical results from finite element analyses presented in [4]. These analyses refer to strip footings resting on a purely elastic half-space. It is assumed that these results satisfactorily describe the response of circular footings on elastic soils and subject to planar loading as well. The results have been obtained by fixing the applied vertical force on the footing and then increasing the applied moment until the toppling of the structure. This means that the increment of the vertical force is zero so from (6), we can write:

$$(11) \quad \dot{Q}_N = K_{NN}\dot{q}_N^{el} + K_{NM}\dot{q}_M^{el} = 0$$

The increment of the moment is written similarly:

$$(12) \quad \dot{Q}_M = K_{MN}\dot{q}_N^{el} + K_{MM}\dot{q}_M^{el}$$

The two main approximation relations introduced in [4] with respect to the numerical results are the following:

$$\text{(13)} \quad \frac{Q_M}{Q_{M,0}} = 2 - \frac{q_{M,0}^{\text{el}}}{q_M^{\text{el}}}, \quad \text{for } |Q_M| > |Q_{M,0}|$$

$$\text{(14)} \quad \frac{\dot{q}_N^{\text{el}}}{\dot{q}_M^{\text{el}}} = -\frac{1}{2}\left(1 - \frac{q_{M,0}^{\text{el}}}{q_M^{\text{el}}}\right)$$

The expression (13) provides the diagram $Q_M - q_M^{\text{el}}$ and the expression (14) yields the coupling between the vertical force and the moment during uplift. These two equations do not suffice for the calculation of the elements of the stiffness matrix, so we introduce two additional assumptions:

***i.*** The elastic stiffness matrix is symmetric (Note that there is no particular reason for it to be symmetric, in general). This is particularly helpful for the numerical treatment of the problem.

***ii.*** The element $K_{NN}$ remains constant during uplift. This means that all the effects of uplift on the vertical force and vertical displacement of footing will by attributed to the coupling term $K_{MN} = K_{NM}$.

The approximation relations and the aforementioned assumptions lead to the following elastic stiffness matrix:

$$\text{(15)} \quad \begin{bmatrix} \dot{Q}_N \\ \dot{Q}_V \\ \dot{Q}_M \end{bmatrix} = \begin{bmatrix} K_{NN} & 0 & K_{NM} \\ 0 & K_{VV} & 0 \\ K_{MN} & 0 & K_{MM} \end{bmatrix} \begin{bmatrix} \dot{q}_N^{\text{el}} \\ \dot{q}_V^{\text{el}} \\ \dot{q}_M^{\text{el}} \end{bmatrix}$$

with:

$$\text{(16)} \quad K_{NN} = \widetilde{K}_{NN}$$

$$\text{(17)} \quad K_{VV} = \widetilde{K}_{VV}$$

$$\text{(18)} \quad K_{NM} = K_{MN} = \begin{cases} 0, & \text{if } |q_M^{\text{el}}| \leq |q_{M,0}^{\text{el}}| \\ \dfrac{1}{2}\widetilde{K}_{NN}\left(1 - \dfrac{q_{M,0}^{\text{el}}}{q_M^{\text{el}}}\right), & \text{if } |q_M^{\text{el}}| > |q_{M,0}^{\text{el}}| \end{cases}$$

$$\text{(19)} \quad K_{MM} = \begin{cases} \widetilde{K}_{MM}, & \text{if } |q_M^{\text{el}}| \leq |q_{M,0}^{\text{el}}| \\ \widetilde{K}_{MM}\left(\dfrac{q_{M,0}^{\text{el}}}{q_M^{\text{el}}}\right)^2 + \dfrac{1}{4}\widetilde{K}_{NN}\left(1 - \dfrac{q_{M,0}^{\text{el}}}{q_M^{\text{el}}}\right)^2, & \text{if } |q_M^{\text{el}}| > |q_{M,0}^{\text{el}}| \end{cases}$$

The quantity $q_{M,0}^{\text{el}}$, which is generally a function of the applied vertical force, is given for an elastic half-space by (10).

## 2.4 Plasticity model

For the description of the mechanism of soil plasticity we develop a "bounding surface" hypoplastic model following the formulation presented in [22]. The advantages of this formulation, which is particularly oriented for the description of cyclic behavior, are its simplicity and flexibility, both of which are particularly desirable for the numerical treatment of the problem and for the investigation of uplift-plasticity coupling. The principal feature of the model is the introduction of a surface in the space of generalized forces, called **bounding surface**, whose main role is the evaluation of the magnitude of the plastic modulus. The bounding surface also serves in the definition of the direction of the plastic displacement increment.

2.4.1 Bounding surface

Following the reasoning of §2.2, the bounding surface of the proposed model is identified with an ellipsoid centred at the origin in the space of the force parameters. It can thus be described by the equation:

$$(20) \quad f_{\text{BS}}\left(\underline{Q}\right) \equiv Q_N^2 + \left(\frac{Q_V}{Q_{V,\max}}\right)^2 + \left(\frac{Q_M}{Q_{M,\max}}\right)^2 = 1$$

We note that in the ***particular case*** where no uplift of the footing is considered, this surface is identified with the surface of ultimate loads of the system and more elaborate approximations could thus be considered. The proposed ellipsoidal bounding surface, while being extremely simple, retains a more than sufficient level of accuracy with respect to the real behaviour. The bounding surface is represented in Figure 7.

The role of the bounding surface is two-fold:

***a.*** To define the cases of pure loading, unloading and neutral loading.

***b.*** To define the direction of the plastic displacement increment and the magnitude of the plastic modulus.

These two goals are achieved by introducing a mapping rule, which maps every point in the interior of the bounding surface to a specific point, called ***image-point***, on the surface boundary. For every point **P** at the interior of the bounding surface we define its corresponding image point using a radial rule as follows (cf. Figure 7):

$$(21) \quad \mathbf{I_P} = \left\{ \lambda \mathbf{P} \mid \mathbf{I_P} \in \partial f_{\text{BS}} \ \text{ et } \ \lambda \geq 1 \right\}$$

Given a current state of the generalized forces $\underline{Q}$ associated with a point **P**, we can identify whether an increment $\underline{\dot{Q}}$ produces a pure loading, neutral loading or unloading response by evaluating the unit normal vector at the image point $\mathbf{I_P}$ :

$$(22) \quad \underline{n} = \left( \left. \frac{\partial f_{\text{BS}}}{\partial \underline{Q}} \right|_{\mathbf{I_P}} \right) \Big/ \left\| \left. \frac{\partial f_{\text{BS}}}{\partial \underline{Q}} \right|_{\mathbf{I_P}} \right\|$$

With direction of the force increment being defined by:

$$(23) \quad \underline{n}_Q = \frac{1}{\left\|\underline{\dot{Q}}\right\|} \underline{\dot{Q}},$$

the cases of pure loading, neutral loading and unloading are defined as follows:

$$\begin{aligned} \underline{n}_Q \cdot \underline{n} > 0 & \quad \text{loading} \\ \underline{n}_Q \cdot \underline{n} = 0 & \quad \text{neutral loading} \\ \underline{n}_Q \cdot \underline{n} < 0 & \quad \text{unloading} \end{aligned} \quad (24)$$

Pure loading is accompanied by the development of plastic displacements. In the cases of neutral loading and unloading, the response is purely elastic.

### 2.4.2 Definition of the plastic modulus

In the case of pure loading, we introduce a generalized plastic modulus $\underline{\underline{\mathcal{H}}}$ by writing:

$$\dot{\underline{Q}} = \underline{\underline{\mathcal{H}}} \dot{\underline{q}}^{\text{pl}} \quad (25)$$

We will assume that an inverse always exist, so we obtain:

$$\dot{\underline{q}}^{\text{pl}} = \underline{\underline{\mathcal{H}}}^{-1} \dot{\underline{Q}} \quad (26)$$

Following the reasoning presented in [23], we can write the inverse of the plastic modulus in the form:

$$\underline{\underline{\mathcal{H}}}^{-1} = \frac{1}{h}(\underline{n}_g \otimes \underline{n}) \quad (27)$$

In (27), $h$ is a scalar function and $\underline{n}$ is defined as in (22). The direction of the plastic displacement increment is then controlled by the unit vector $\underline{n}_g$. If:

$$\underline{n}_g \equiv \underline{n}, \quad (28)$$

the model is associated. If not, the vector $\underline{n}_g$ may be defined by introduction of a plastic potential.

The magnitude of the plastic displacement increment is controlled by the scalar quantity $h$. In the context of bounding surface plasticity, this quantity is defined as a function of the distance between the current state of forces and its image point. A simple measure of this distance is given by the scalar $\lambda$ in equation (21). We can thus write:

$$h = h(\lambda) \quad (29)$$

The expression (29) can be calibrated using numerical or experimental results. If the loading of the footing under a concentric vertical force is considered, a logarithmic variation of the plastic modulus (*cf.* [24]) may be adopted, leading to the particularly simple expression:

$$h = h_0 \ln(\lambda) \quad (30)$$

with $h_0$ being a numerical parameter.

A more complicated formulation may be introduced to take into account the history of loading of the system. For example, in [15] the bounding surface evolves following an isotropic hardening rule in pure loading, whereas in unloading/reloading the bounding surface remains fixed and the plastic modulus is defined via the image point of the current state of forces as explained above. In the present formulation, a simpler account of the loading history will be adopted by writing:

(31) $$h = h_0 \ln\left[\left(\frac{\lambda}{\lambda_{\min}}\right)^{p_1} \lambda\right]$$

where $\lambda_{\min}$ is the minimum value of $\lambda$ obtained during loading and $p_1$ a numerical parameter. The meaning of (30) and (31) is the following: for pure loading $\lambda_{\min} = \lambda$ and we recuperate (30). If $\lambda$ is large, $h$ is also large and the magnitude of plastic displacement increment is small so the response is principally elastic. On the contrary, for $\lambda$ small, $h$ is also small and the plastic displacement increment is large. In the case where the state of forces reaches the bounding surface $\lambda \to 1$, thus $h \to 0$ and the system is led to a state of plastic flow. In the phase of reloading $\lambda_{\min} < \lambda$ and the response of the system is less plastic than in the phase of pure loading.

## 2.5 Model parameters and uplift-plasticity coupling

In this section, we summarize the parameters of the model and we comment on their determination. We also comment on the uplift-plasticity coupling within the macroelement. Concerning the model parameters, we have:

- $D$: the only geometrical parameter is the footing diameter, which is prescribed.

- $N_{\max}$: the maximum vertical force supported by the footing may be calculated as follows. If the soil exhibits a uniform soil cohesion $C_0$, we obtain (*cf.* [25]):

(32) $$N_{\max} = 6.06 C_0 \frac{\pi D^2}{4},$$

For heterogeneous soil conditions, the solutions by Salençon & Matar (*cf.* [26]) may be used. They provide, among others, the maximum concentric vertical force for circular footings on soils exhibiting a cohesion varying linearly with depth.

- *Static impedances*: the commonly used expressions for circular footings on an elastic half-space with constant shear modulus $\mathcal{G}$ and Poisson's ratio $\nu$ are recalled (cf. [27]). Following the normalization scheme (4), they are written as follows:

(33) $$\widetilde{K}_{NN} = \frac{2\mathcal{G}D^2}{N_{\max}(1-\nu)}$$

(34) $$\widetilde{K}_{VV} = \frac{4\mathcal{G}D^2}{N_{\max}(2-\nu)}$$

(35) $$\widetilde{K}_{MM} = \frac{\mathcal{G}D^2}{3N_{\max}(1-\nu)}$$

- *Bounding surface*: Concerning the parameters defining the bounding surface, we have:

(36) $$Q_{V,\max} = \frac{C_0 \pi D^2}{4 N_{\max}}$$

In (36), $C_0$ designates the soil cohesion at the surface. For the determination of $Q_{M,\max}$ we can use the upper bounds established in [19]. These give for a homogeneous soil:

$$Q_{M,\max} = \frac{0.67 C_0 \pi D^2}{4 N_{\max}} \quad (37)$$

- *Plastic modulus*: the parameters $h_0$ and $p_1$ describe the evolution of the magnitude of plastic modulus and may vary significantly for different soil formations. It is thus rather ambiguous to propose prescriptions for these parameters. In lack of numerical or experimental results pertaining to a specific soil, we will adopt the following characteristic values and we will limit ourselves to a qualitative description of the system behaviour:

$$h_0 = 0.1 \widetilde{K}_{NN}$$

$$p_1 = 5$$

We may also note that the relation (31) is not restrictive. It can be improved or replaced by any other relation approximating the variation of the scalar quantity $h$ implying the introduction of additional parameters $p_2, p_3, \ldots$.

- *Uplift-plasticity coupling*: the uplift-plasticity coupling is introduced in the first place by the relation which provides the moment (or equivalently the elastic rotation angle) of uplift initiation as a function of the applied vertical force. This relation for an elastic soil is linear as in (10). For an elastoplastic soil, the relation (10) may be replaced by an approximation relation of the following form, as has been proposed by Crémer (*cf.* [4]):

$$q_{M,0}^{\text{el}} = \pm \frac{1}{d_1} \left( \frac{Q_N}{\widetilde{K}_{MM}} \right) e^{-d_2 Q_N} \quad (38)$$

For strip footings, Crémer (*cf.* [4]) proposes:

$$d_1 = 4, \quad d_2 = 2.5$$

For circular footings, we may adopt:

$$d_1 = 3, \quad d_2 = 2$$

As it was the case for the approximation relation of the scalar quantity $h$, the relation (38) may be improved or replaced with respect to specific numerical or experimental results (implying the introduction of uplift parameters $d_1, d_2, \ldots$). Besides relation (38), the uplift-plasticity coupling within the macroelement may be directly obtained without the introduction of additional parameters by observing that the vertical force $Q_N$ is varying as a plastic response is obtained. Consequently, the entire non-linear elastic tangent stiffness matrix is varying since its elements are functions of $q_{M,0}^{\text{el}}$ which in turn is a function of $Q_N$.

## 3 Behavior under quasistatic loading

We investigate in this section the system response under quasistatic monotonic or cyclic loading. This is achieved by performing numerical displacement-controlled loading tests. A specific displacement history is prescribed for the footing center and the force-displacement response of the system is recorded. Since no strict calibration of the model parameters has been performed, we will limit ourselves in investigating the qualitative aspects of the system response.

## 3.1 Vertical force – vertical displacement

We initially examine the system response under a prescribed history of vertical displacement $q_N$. The results are presented in Figure 8(a) providing the model purely elastic and fully elastoplastic response in cycles of loading-unloading and in Figure 8(b) presenting corresponding experimental results obtained in [28]. The bounding surface hypoplastic model predicts a smooth transition towards plastic flow and between the phases of reloading and pure loading. This is a feature that agrees well with the observed soil behavior. We also note that the elastic response of the system (*cf.* Figure 8(b) in unloading) is almost negligible. This can be achieved in the model, by prescribing a plastic modulus parameter $h_0$ considerably smaller than $\widetilde{K}_{NN}$.

## 3.2 Horizontal force – horizontal displacement

We examine next the model response under a loading in the horizontal direction. The numerical tests performed comprise two stages:

***a.*** Initially, a prescribed vertical displacement is applied that activates the plasticity mechanism up to a certain level.

***b.*** The vertical position of the footing being kept fixed, a prescribed horizontal displacement history is then applied. The horizontal force-horizontal displacement diagram is recorded as well as the trace of the current state of forces in the space of the generalized force parameters.

The results of two such tests are presented in Figure 9. In Figure 9(a) the results from a test under monotonic horizontal loading are presented. A vertical displacement $q_N = 0.03$ is initially applied and then a horizontal displacement equal to $q_V = 0.01$. In the diagram $Q_V - q_V$ both the purely elastic and the fully elastoplastic response of the system are presented. The diagram reveals a non-linear behaviour from the beginning of the loading and a smooth transition towards failure. The trace of the force state in the space of the generalized force parameters (plane $Q_V - Q_N$ in Figure 9) shows a considerable decrease of the vertical force as the horizontal displacement is increased. The trace of the force follows the elliptical shape of the bounding surface (also plotted in the diagram $Q_V - Q_N$) and approaches to it asymptotically. This coupling between the horizontal and the vertical force is an essential feature of the elastoplastic response of the system that has been verified experimentally as in [28]. Such a coupling can by no means be captured by decoupled springs in the horizontal and vertical direction, even if they exhibit an advanced elastoplastic constitutive law.

In Figure 9(b), the results of a test under horizontal cyclic loading are presented. A vertical displacement $q_N = 0.01$ is initially applied and then five successive cycles of horizontal displacement with amplitude increasing from $q_V = \pm 0.001$ to $q_V = \pm 0.005$ with a step of 0.001 are imposed on the footing. The $Q_V - q_V$ diagram reveals the dissipative nature of the plasticity mechanism, with cycles of energy dissipation increasing with increasing amplitude of horizontal displacement. On the contrary, the elastic behaviour of the system is fully reversible. The diagram $Q_V - Q_N$ presents a « saturation » of the response with the accumulation of cycles towards the origin. This means that the plastic response becomes less and less pronounced and it is recuperated only when the trace of the forces approaches the bounding surface. Even so, the transition between the phases of reloading and pure loading remains always smooth. In contrast, the transition between unloading and reloading is not smooth as it is also shown in the diagram $Q_V - Q_N$. This is a deliberate option, so that the system response during unloading is purely elastic. However, if so desired, it is possible for the bounding surface hypoplastic model to accommodate for a smooth transition even between the phases of unloading and reloading.

## 3.3 Moment – rotation angle

The response of the system under moment loading is investigated next. The stage **b** of the numerical tests performed for horizontal loading is here replaced by the application of a prescribed history of the rotation angle of the footing, its vertical position being kept fixed. Four tests under monotonic loading are presented in Figure 10. In these tests, the following displacement histories are applied:

(a) :   Vertical displacement $q_N = 0.03$, then rotation angle $q_M = 0.003$.

(b) :   Vertical displacement $q_N = 0.01$, then rotation angle $q_M = 0.003$.

(c) :   Vertical displacement $q_N = 0.005$, then rotation angle $q_M = 0.003$.

(d) :   Vertical displacement $q_N = 0.0005$, then rotation angle $q_M = 0.003$.

For every test, we present two diagrams: the diagram $Q_M - q_M$ and the trace of the vector $Q$ in the space of the generalized forces (plane $Q_M - Q_N$). This second diagram contains three additional curves: the elliptical bounding surface, the curve which corresponds to the initiation of uplift and finally the surface of ultimate loads of the system (rugby ball-shaped curve). The curve of uplift initiation is given by the expression:

$$(39) \qquad Q_{M,0} = \pm \frac{Q_N}{3} e^{-2Q_N}$$

The equation for the surface of ultimate loads is the one proposed in [4]:

$$(40) \qquad Q_M = \pm 0.37 Q_N^{0.8} \left(1 - Q_N\right)^{0.8}$$

The $Q_M - q_M$ diagrams show that the response of the system is no longer linear, once the uplift initiation curve is surpassed. In the $Q_M - Q_N$ diagrams, the results reveal how the mechanisms of plasticity and uplift are combined to provide the admissible states of forces of the system that should be included in the interior of the surface of ultimate loads. As the imposed vertical displacement becomes smaller and smaller, the effect of the plasticity mechanism is gradually decreased. In parallel, the effect of uplift becomes more and more important and, once the uplift initiation curve is surpassed, the trace of the forces in the plane $Q_M - Q_N$, instead of following the elliptical shape of the bounding surface, changes direction and follows the shape of the surface of ultimate loads. The results show that although the surface of ultimate loads is not explicitly used in the formulation of neither the plasticity nor the uplift model, it can be obtained as a combined result of the two. It is thus possible to formulate both models independently (respecting their particular characteristics), but in such a way so that the desired ultimate surface is eventually obtained.

As far as cyclic response is concerned, Figure 11 presents the diagrams $Q_N - q_N$, $Q_M - q_M$ and $Q_M - Q_N$ for two tests :

(a) :   Vertical displacement $q_N = 0.001$, then one single cycle of rotation angle $q_M = \pm 0.001$.

(b) :   Vertical displacement $q_N = 0.03$, then five successive cycles of increasing rotation angle from $q_M = \pm 0.0005$ to $q_M = \pm 0.0025$ with a step of $0.0005$.

The test (a), performed for a small vertical displacement, is mainly governed by the uplift mechanism. It reveals the non-linear elastic behaviour during uplift (curve of elastic response in diagram $Q_M - q_M$) with a cycle of energy dissipation obtained from the plasticity mechanism and the moment-vertical force coupling as shown by the increase in $Q_N$ as the footing center tends to be lifted but the imposed vertical displacement obliges it to stay fixed with respect to its vertical position (diagram $Q_N - q_N$). The results of test (b) clearly present how the mechanism of uplift becomes more and more important as the cycles of increasing angle of rotation are accumulated. As the trace of the current force state is dragged towards the origin and the imposed rotation angle is increased, the change in the direction of the trace of current force due to uplift becomes more and more pronounced. This change in direction depicts the required increase in the vertical force (as in test (a)) that has to be applied in the footing to keep it in a fixed vertical position, as the imposed rotation angle increase forces it to uplift. The diagram $Q_M - q_M$ of test (b) clearly presents the obtained *S*-form of the moment-rotation angle curve indicating the reduction of the apparent rotational stiffness of the footing due to the initiation of the uplift mechanism.

## 4  Extension to dynamic loading

The structure of the macroelement model presented so far refers to the system behavior under quasistatic monotonic or cyclic loading (terms in the equilibrium equations associated with acceleration and velocity have been neglected). However, the principal domain of application for the macroelement is implementing efficient non-linear dynamic soil-structure interaction analyses. In this paragraph, we explain under which assumptions the macroelement is incorporated into the global superstructure model and we present, in a numerical application of the dynamic analysis of a real structure, the type of results that can be obtained with the macroelement.

### 4.1  General principles

The extension of the domain of application of the macroelement to dynamic loading conditions is performed by considering that the soil domain is divided into two separate sub-domains: the near field and the far field. The near field is identified as the soil sub-domain at the footing vicinity where all the non-linearities (material and geometric) of the system take place. These are described within the macroelement by the plasticity and the uplift model as it has been explained. The far field, on the other hand, is the soil sub-domain where the response remains purely linear. This distinction between near and far fields allows describing the contribution of the far field on the system behavior using the dynamic (elastic) impedances of the foundation and the contribution of the near field with the macroelement. Extension to dynamic loading conditions may thus be achieved by performing the following:

*a.* Identifying the parameters $\widetilde{K}_{NN}, \widetilde{K}_{VV}, \widetilde{K}_{MM}$ in the uplift non-linear elastic model of the macroelement as the real part of the corresponding dynamic impedances of the footing. If the dynamic impedances of a circular footing resting on the soil surface are to be used, this actually implies that the near field is actually reduced to the plane directly below the footing. We also note that in the present state of macroelement development the resolution of the system is performed in the time domain and no dependence of the dynamic impedances on the frequency of excitation is considered. The retained dynamic impedances can thus correspond to some characteristic frequency of the system, such as its fundamental eigen-frequency *etc*.

***b.*** Introducing the imaginary part of the retained dynamic impedances in order to account for the phenomenon of radiation damping.

***c.*** There is no need to account for material damping as this is actually reproduced by the plasticity model of the macroelement.

## 4.2 Numerical application

In order to show the type of results that can be obtained using the macroelement we present an example of application of the proposed model for the dynamic analysis of a real structure: a column of the Arc Viaduct (France) subject to a real seismic acceleration time history. The examined structure is presented in Figure 12(a) and is modelled with a simple structural model as the one presented in Figure 12(b). The model exhibits four degrees of freedom: the horizontal translation of the Viaduct deck as well as the horizontal and vertical translation and the rotation of the column foundation. The latter three will be described by the macroelement.

For the required model parameters we introduce the following values:

- Mass of the superstructure : $m_S = 1.5 \times 10^6 \, [\text{kgr}]$
- Mass of the foundation : $m_F = 0.5 \times 10^6 \, [\text{kgr}]$
- Mass moment of inertia of the foundation : $J_F = 22.1 \times 10^6 \, [\text{kgr} \times \text{m}^2]$
- Height of the superstructure : $H = 15 \, [\text{m}]$
- Percentage of effective damping : $\xi_S = 7\%$

The elastic properties of the column are:

- Cross-sectional are of the column : $A = 12.9 \, [\text{m}^2]$
- Young's modulus : $E = 35 \, [\text{GPa}]$
- Second moment of area : $I = 20.2 \, [\text{m}^4]$

The foundation of the column comprises a rigid circular footing resting on the surface of a homogeneous purely cohesive soil. The following are defined:

- Footing diameter : $D = 12 \, [\text{m}]$
- Soil uniform cohesion : $C_0 = 50 \, [\text{kPa}]$

The bearing capacity safety factor against a concentric vertical load is $FS = 1.75$

Concerning the foundation impedances, we will use the approximate relations proposed in [27] which are presented in *Table 2*. We note that the elastic shear modulus is given as a function of the shear wave velocity $V_s$ and the mass density $\rho$ of the soil by the following relationship:

(41) $$\mathcal{G} = V_s^2 \rho$$

We also calculate the velocity of Lysmer's analogue, necessary for the evaluation of the radiation damping coefficients. This quantity is given by the expression:

$$(42) \quad V_{\text{La}} = \frac{3.4}{\pi(1-\nu)} V_S$$

As a first approximation, we will consider that the foundation dynamic impedances are frequency-independent and that their real part is equal to the static impedances of the footing. We will consider a shear wave velocity $V_s = 200 \,[\text{m/sec}]$, Poisson's ratio $\nu = 0.5$ (undrained conditions) and $\rho = 2000 \,[\text{kgr/m}^3]$. These values lead to the following values for the stiffness and radiation damping coefficients:

- $\widetilde{K}_{NN} = 3840000 \,[\text{kN/m}]$
- $\widetilde{K}_{VV} = 2560000 \,[\text{kN/m}]$
- $\widetilde{K}_{MM} = 92160000 \,[\text{kNm/rad}]$
- $C_{NN} = 97920 \,[\text{kNs/m}]$
- $C_{VV} = 45239 \,[\text{kNs/m}]$
- $C_{MM} = 881280 \,[\text{kNms/rad}]$

We will consider the response of the structure subject to the acceleration time history recorded during the Friuli earthquake (Italy, 1976) which is represented in Figure 13.

The recorded maximum horizontal acceleration is $a_{\max} = 2.5 \,[\text{m/sec}^2] = 0.25g$. We consider that the acceleration acts at the horizontal direction. The vertical component of the input motion is zero. In the following, we present the results of the dynamic analysis for three different cases:

***a.*** Linear elastic behaviour (Uplift and plasticity mechanisms deactivated).

***b.*** Elastic behaviour with uplift (Plasticity mechanism deactivated).

***c.*** Fully elastoplastic behaviour with uplift.

4.2.1 Linear elastic behaviour

The results are presented in Figure 14. The figure contains nine diagrams:

- The three time histories of horizontal force, moment and vertical force acting on the footing.

- The three time histories of the horizontal displacement, rotation angle and vertical displacement at the center of the footing.

- The horizontal force – horizontal displacement, moment – rotation angle and vertical force – vertical displacement diagrams.

We note that the maximum recorded moment is $M_{\max} = 5 \times 10^7 \,[\text{Nm}]$ and the maximum horizontal force $V_{\max} = 3.2 \times 10^6 \,[\text{N}]$. The force – displacement diagrams present indeed a purely linear response and no coupling with the vertical force is obtained.

4.2.2 Elastic behaviour with uplift

In Figure 15 we present the same diagrams as before, but now we activate the mechanism of uplift within the macroelement.

It is interesting to note the following:

*a.* While the horizontal force – horizontal displacement diagram remains purely linear, the moment – rotation angle diagram exhibits the characteristic *S*-shaped form due to uplift inducing an apparent reduction in the rotational stiffness of the footing. We note that the behaviour is reversible with almost zero energy dissipation (some dissipation exists however due to radiation damping).

*b.* The diagrams related to the vertical force show the coupling that is obtained during uplift between the vertical force and the moment. The behaviour is always reversible. The small residual displacement that seems to be obtained in the vertical displacement time history is an accumulated numerical error due to the algorithmic treatment of the uplift model, in which the elastic tangent stiffness matrix is approximated with a first-order Taylor approximation in a purely explicit manner. This should be a point of improvement in future updates of the proposed model. On the contrary, no coupling is obtained during uplift with respect to the horizontal force.

*c.* The activation of uplift mechanism leads to a reduction in the maximum recorded moment, which is now $M_{\max} = 4 \times 10^7 \, [\text{Nm}]$. In this sense, the uplift mechanism acts as a mechanism of seismic isolation for the superstructure. The maximum recorded horizontal force is not affected by the activation of the uplift mechanism. It is equal to $V_{\max} = 3.2 \times 10^6 \, [\text{N}]$.

### 4.2.3 Fully elastoplastic response with uplift

Finally, in Figure 16 we present the analysis results considering fully elastoplastic response of the soil with uplift. In the macroelement, both the plasticity and the uplift mechanisms are activated.

Principal conclusions drawn from the results are:

*a.* The development of a residual horizontal displacement. Its magnitude is not significant, but it shows that the model can predict the development of residual displacements/rotations at the foundation level.

*b.* The development of cycles of energy dissipation in the horizontal force – horizontal displacement and moment – rotation angle diagrams. It is interesting to note the difference in the form of the cycles between the two diagrams. The absence of uplift in the horizontal force gives rise to a more regular form for the obtained cycles of energy dissipation.

*c.* The accumulated vertical displacement (settlement) of the footing during loading. This is obtained as the plasticity model couples all three degrees-of-freedom of the foundation.

*d.* The significant reduction in both the maximum recorded moment ($M_{\max} = 1.5 \times 10^7 \, [\text{Nm}]$) and the maximum recorded horizontal force ($V_{\max} = 1.3 \times 10^6 \, [\text{N}]$) acting at the center of the footing.

## 5  Conclusions

We have presented in this article a macroelement model for shallow foundations intended to serve as a practical tool for efficient non-linear dynamic soil-structure interaction analyses. The model comprises both the geometric and the material non-linearity of the system in a simple and coherent way that respects the particular characteristics of each mechanism. In particular, it incorporates an associated plasticity model that can accounts for the soil elastoplastic response in undrained conditions and a phenomenological non-linear elastic model for the uplift mechanism, respecting its reversible and non-dissipative nature. It also allows recuperating the surface of ultimate loads of the system as the combined result of both the uplift and the plasticity mechanisms, attributing to the ultimate surface its real meaning from the point of view of Yield Design theory (*cf.* [29]): domain of all the combinations of loads that can be supported by the system. Moreover, it highlights the importance of clearly defining the strength and resistance properties of both the soil and the soil-footing interface.

From the point of view of its numerical treatment, the proposed formulation induces a particularly simple rheological model and allows for a simultaneous resolution of both mechanisms. It is particularly flexible in modifying, activating or deactivating different mechanisms, since the modeling of each mechanism is performed independently. Moreover, its extension to fully three-dimensional configurations is straightforward, especially in what regards the uplift model that is written with respect to the generalized displacement parameters without introduction of additional parameters linked with the footing geometry.

However, the results presented have only offered a validation of the qualitative aspects of the system response. It is thus essential that the model parameters be calibrated using numerical or experimental results specifically conceived for the needs of the model. This is particularly necessary for the uplift model, which requires a series of numerical results of the uplift of a circular footing that rests on an elastic half-space.

Possible additional improvements of the model would include among others:

*a.* An implicit resolution scheme for the uplift non-linear elastic model.

*b.* The introduction of the effects of the seismic acceleration on the bearing capacity of the foundation. This has been the subject of recent work by the authors (cf. [18]) and it can be performed by introducing the variation of $N_{max}$ due to the incident acceleration in each time step.

*c.* An extension of the model in the case of a frictional material obeying to the Mohr-Coulomb strength criterion, which can be incorporated within a non-associated plasticity model. It is deemed that the main structure of the macroelement can be preserved, thus giving rise to a generic macroelement structure. Modifications in the plasticity model will only need to be introduced.

*d.* The consideration of the dependence of the dynamic impedances of the foundation on the frequency of excitation.

# 6 Acknowledgments

The first author wishes to thank the École Polytechnique and the Public Benefit Foundation "Alexandros S. Onassis" for the financial support during the execution of this study.

# 7 List of main symbols

**Latin**

| | |
|---|---|
| $C_{ij}$, $i,j = N,V,M$ | *Imaginary part of the dynamic impedances of the foundation (radiation damping)* |
| $C_0$ | *Uniform soil cohesion, soil cohesion at the surface of the ground* |
| $d_1,\ldots$ | *Parameters of uplift model* |
| $D$ | *Diameter of circular footing* |
| $E$ | *Young's modulus of elasticity* |
| $f_{\text{BS}}(\underline{Q})$ | *Analytical expression of the bounding surface in the space of generalized forces* |
| $FS$ | *Bearing capacity safety factor against concentric vertical loading* |
| $\mathcal{G}$ | *Elastic shear modulus* |
| $h$ | *Scalar quantity used for the definition of the plastic modulus* |
| $h_0$ | *Numerical parameter used for the definition of the scalar quantity $h$* |
| $H$ | *Height (of a superstructure)* |
| $\underline{\underline{\mathcal{H}}}$ | *Generalized plastic modulus* |
| $I$ | *Second moment of area* |
| $\mathbf{I}(\mathbf{P})$ | *Image point on the bounding surface of a point $\mathbf{P}$* |
| $J$ | *Mass moment of inertia* |
| $K_{ij}$, $i,j = N,V,M$ | *Normalized elements of the elastic stiffness matrix* |
| $\widetilde{K}_{ij}$, $i,j = N,V,M$ | *Normalized static impedances of the foundation or real part of dynamic impedances* |
| $\mathbf{K}_{ij}$, $i,j = N,V,M$ | *Dimensional elements of the elastic stiffness matrix* |
| $\underline{\underline{\mathcal{K}}}$ | *Elastic stiffness matrix* |
| $m$ | *Mass* |
| $M$ | *Moment applied on the footing* |
| $N$ | *Vertical force applied on the footing* |
| $N_{\max}$ | *Maximum vertical force supported by the footing* |
| $\underline{n}$ | *Unit normal vector on the bounding surface* |
| $\underline{n}_Q$ | *Unit normal vector following the direction of the force increment* |
| $p_1,\ldots$ | *Parameters of the plasticity model* |
| $q_i$, $i = N,V,M$ | *Normalized cinematic parameters of the macroelement* |
| $\underline{q}$ | *Vector of normalized cinematic parameters of the macroelement* |
| $\underline{q}^{\text{el}}, \underline{q}^{\text{pl}}$ | *Elastic and plastic parts of the vector of normalized cinematic parameters* |
| $q^{\text{el}}_{M,0}$ | *Elastic normalized rotation angle at the moment of uplift initiation* |
| $Q_i$, $i = N,V,M$ | *Normalized force parameters of the macroelement* |
| $\underline{Q}$ | *Vector of normalized force parameters of the macroelement* |
| $Q_{M,0}$ | *Uplift initiation normalized moment* |
| $Q_{V,\max}, Q_{M,\max}$ | *Parameters used for the definition of the bounding surface* |
| $V$ | *Horizontal force on the footing* |
| $\underline{u}, \underline{\dot{u}}, \underline{\ddot{u}}$ | *Displacement, velocity and acceleration field* |
| $\mathcal{W}$ | *Normalized work of external forces* |

| W | *Dimensional work of external forces* |
| $x, y, z$ | *Cartesian coordinates* |

**Greek**

| $\theta$ | *Rotation angle* |
| $\lambda$ | *Measure of the distance between current stress point and its image point* |
| $\lambda_{\min}$ | *Minimum attained value for the quantity $\lambda$ during the loading history* |
| $\nu$ | *Poisson's ratio* |
| $\xi$ | *Percentage of effective damping* |

# 8  References

[1]  Nova R, Montrasio L. Settlements of shallow foundations on sand. *Géotechnique* 1999; **41**(2): 243 – 256.

[2]  Paolucci R. Simplified evaluation of earthquake-induced permanent displacements of shallow foundations. *Journal of Earthquake Engineering* 1997; **1**(3): 563 – 579.

[3]  Pedretti S. *Non linear seismic soil foundation interaction: analysis and modelling method.* PhD Thesis, Dpt Ing Strutturale, Politecnico di Milano, 1998.

[4]  Crémer C. *Modélisation du comportement non linéaire des fondations superficielles sous séisme.* PhD thesis, Laboratoire de Mécanique et de Technologie, ENS – Cachan, France, 2001.

[5]  Crémer C, Pecker A, Davenne L. Cyclic macro-element for soil-structure interaction: material and geometrical non linearities. *International Journal for Numerical and Analytical Methods in Geomechanics* 2001; **25:** 1257 – 1284.

[6]  Crémer C, Pecker A, Davenne L. Modelling of nonlinear dynamic behavior of a shallow strip foundation with macroelement. *Journal of Earthquake Engineering* 2002; **6**(2): 175 – 211.

[7]  Le Pape Y, Sieffert JG, Harlicot P. Analyse non linéaire par macro-éléments du comportement des fondations superficielles sous action sismique. *5ème Colloque National AFPS,* Cachan, France, 19 – 21 Octobre 1999: 207 – 214.

[8]  Le Pape Y, Sieffert JG. Application of thermodynamics to the global modelling of shallow foundations on frictional material. *International Journal for Numerical and Analytical Methods in Geomechanics* 2001; **25**: 1377 – 140

[9]  Houlsby GT, Cassidy MJ, Einav I. A generalised Winkler model for the behaviour of shallow foundations. *Géotechnique* 2005 ; **55**(6): pp. 449 – 460.

[10]  Einav I, Cassidy MJ. A framework for modelling rigid footing behaviour based on energy principles. *Computers and Geotechnics* 2005; **32**: 491 – 504.

[11]  Wood DM, Kalasin T. Macroelement for study of the dynamic response of gravity retaining walls *Cyclic behaviour of soils and liquefaction phenomena,* Triandafyllides N, Ed, Taylor and Francis Group, London, 2004; 551 – 561.

[12]  Gottardi G, Houlsby GT, Butterfield R. Plastic response of circular footings on sand under general planar loading. *Géotechnique* 1999; **49**(4): 453 – 469.

[13]  Martin CM, Houlsby GT. Combined Loading of spudcan foundations on clay: Laboratory tests. *Géotechnique* 2000; **50**(4): 325 – 338.

[14]  Houlsby GT, Cassidy MJ. A plasticity model for the behaviour of footings on sand under combined loading. *Géotechnique* 2002; **52**(2): 117 – 129.

[15]  Di Prisco C, Nova R, Sibilia A. Shallow footing under cyclic loading: Experimental behavior and constitutive modeling. In *Geotechnical analysis of the Seismic Vulnerability of Historical Monuments.* Maugeri M, Nova R, Eds. Pàtron, 2003: 99 – 121.

[16]  Cassidy MJ, Martin CM, Houlsby GT. Development and application of force resultant models describing jack-up foundation behaviour. *Marine Structures* 2004; **17**: 165 – 193.

[17]  Grange S, Kotronis P, Mazars J. Advancement of simplified modeling strategies for 3D phenomena and or/ boundary conditions for base-isolated buildings or specific soil-structure interactions. *European Program LESSLOSS: Risk mitigation for earthquakes and landslides integrated project. Deliverable report 67,* September 2006.

[18]  Chatzigogos CT, Pecker A, Salençon J. Seismic bearing capacity of a circular footing on a heterogeneous cohesive soil, *Soils and Foundations* 2007; accepted for publication.

[19]  Randolph MF, Puzrin AM. Upper bound limit analysis of circular foundations on clay under general loading. *Geotechnique* 2003; **53**(9): 785 – 796.

[20]  Bell RW. The analysis of offshore foundations subjected to combined loading. *MSc thesis, University of Oxford*, 2001.

[21]  Wolf JP. *Soil-structure interaction analysis in the time domain.* Prentice Hall Inc., New Jersey, 1988.

[22]  Dafalias YF, Hermann LR. Bounding surface formulation of soil plasticity. *Soil Mechanics – transient and cyclic loading*, Pande GN, Zienkiewicz OC, Eds, Wiley, 1982: 173 – 218.

[23]  Pastor M, Zienkiewicz OC, Chan AHC. Generalized plasticity and the modeling of soil behavior. *International Journal for Numerical and Analytical Methods in Geomechanics* 1990; 151 – 190.

[24]  Butterfield R. A simple analysis of the load capacity of rigid footings on granular materials. *Journée de Géotechnique* 1980; pp. 128 – 134.

[25]  Eason G, Shield RT. The plastic indentation of a semi – infinite solid by a perfectly rough circular punch. *J. Appl. Math. Phys. (ZAMP)* 1960; **11**(1): pp. 33 – 43.

[26]  Salençon J, Matar M. Capacité portante des fondations superficielles circulaires. *Journal de Mécanique théorique et appliquée* 1982; **1**(2): 237 – 267.

[27]  Sieffert JG, Cevear F. *Manuel des functions d'impédances – Fondations superficielles.* Ouest ed. Presses Académiques, 1992.

[28]  Martin CM, Houlsby GT. Combined loading of spudcan foundations on clay: numerical modeling. *Géotechnique* 2001; **51**(8): 687 - 699.

[29]  Salençon J. An introduction to the Yield Design Theory and its applications in Soil Mechanics. *Eur. J. Mech. A/Solids* 1990; **9**(5): 477 – 500.

# List of Figures

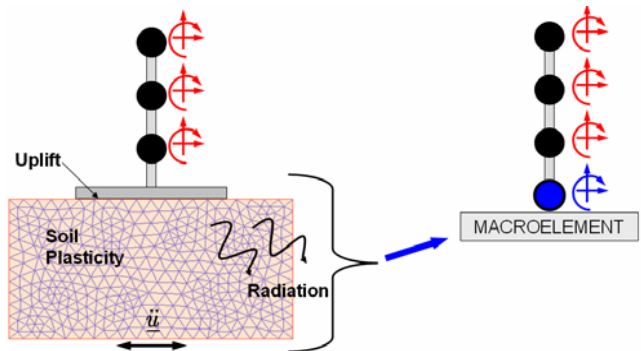

*Figure 1 – Generic soil-foundation-structure system subject to dynamic loading and macroelement concept.*

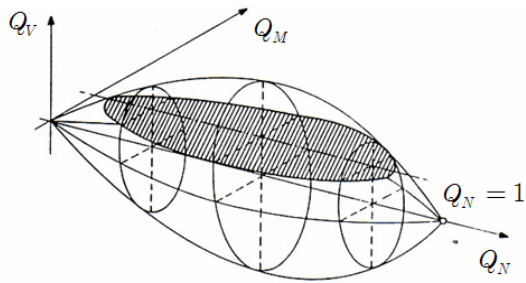

*Figure 2 – Rugby ball shaped surface of ultimate loads identified as the yield surface of the plasticity model in the model of Nova & Montrasio (cf. [1]) and its evolution models.*

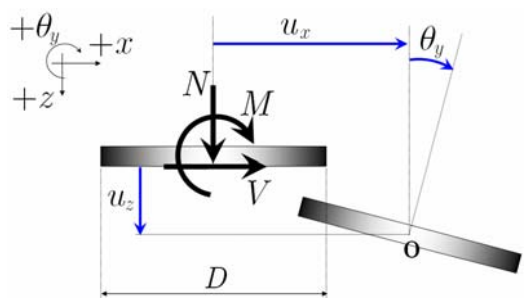

*Figure 3 – Definition of generalized forces and displacements for a perfectly rigid circular footing under planar loading.*

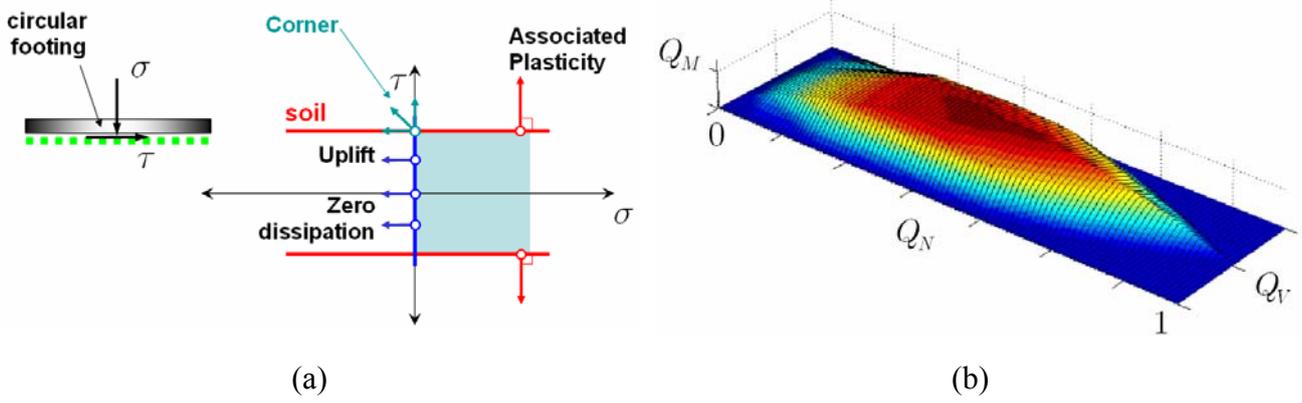

(a)                                                      (b)

*Figure 4 – (a) Assumed behavior at the scale of the constituent materials of the system for a macroelement oriented towards earthquake engineering applications. (b) Corresponding surface of ultimate loads of the system in the space $(Q_N, Q_V, Q_M)$.*

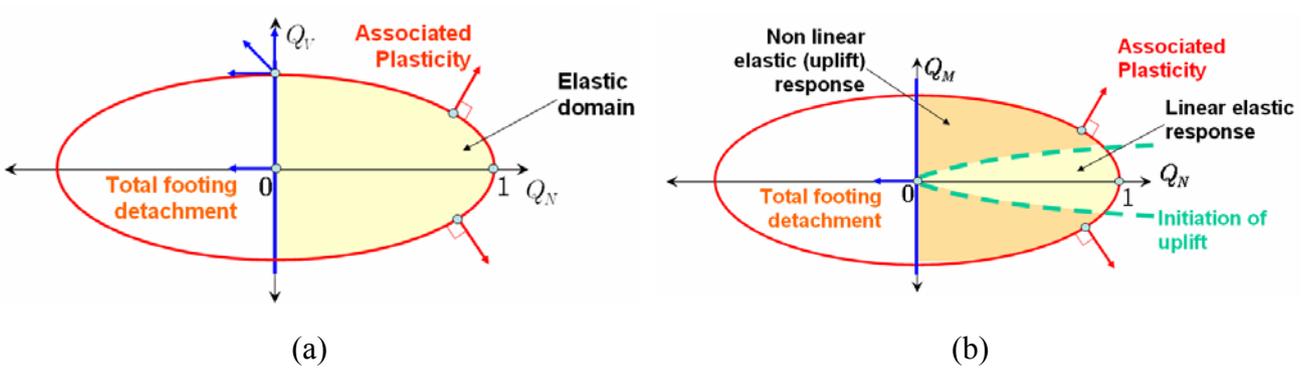

(a)                                                      (b)

*Figure 5 – (a) Structure of the proposed macroelement in the $Q_N - Q_V$ plane and (b) Structure of the proposed macroelement in the $Q_N - Q_M$ plane.*

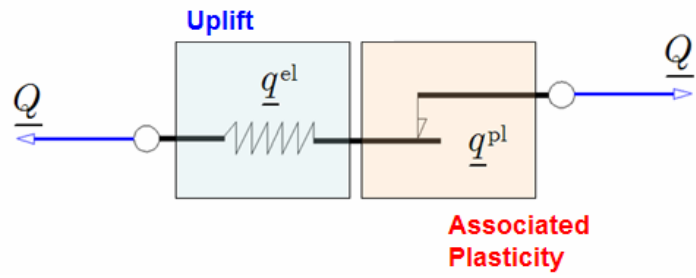
*Figure 6 – Structure of the proposed macroelement*

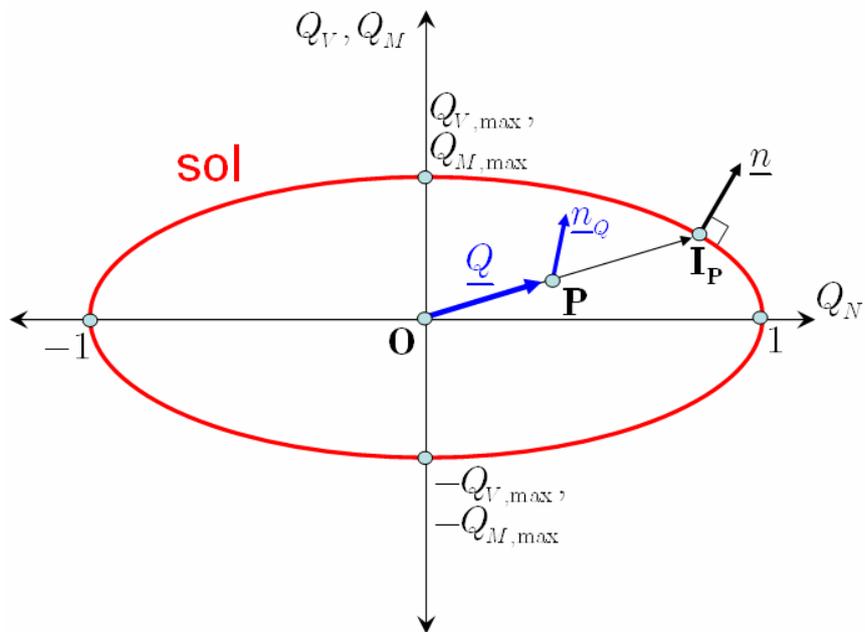
*Figure 7– Surface bornant pour le modèle hypoplastique incorporé dans le macroélément*

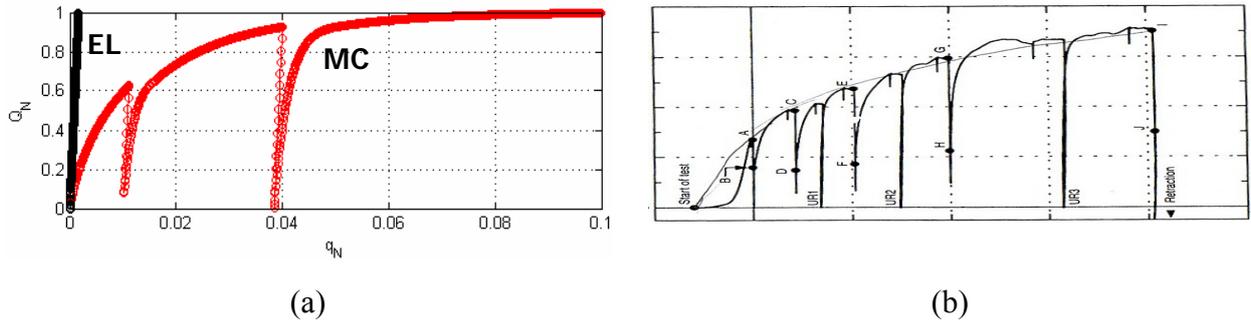

**EL**: *elastic response,* **MC**: *full macroelement response*

*Figure 8 – (a) System response under quasistatic vertical loading and (b) Experimental results obtained in* [28]

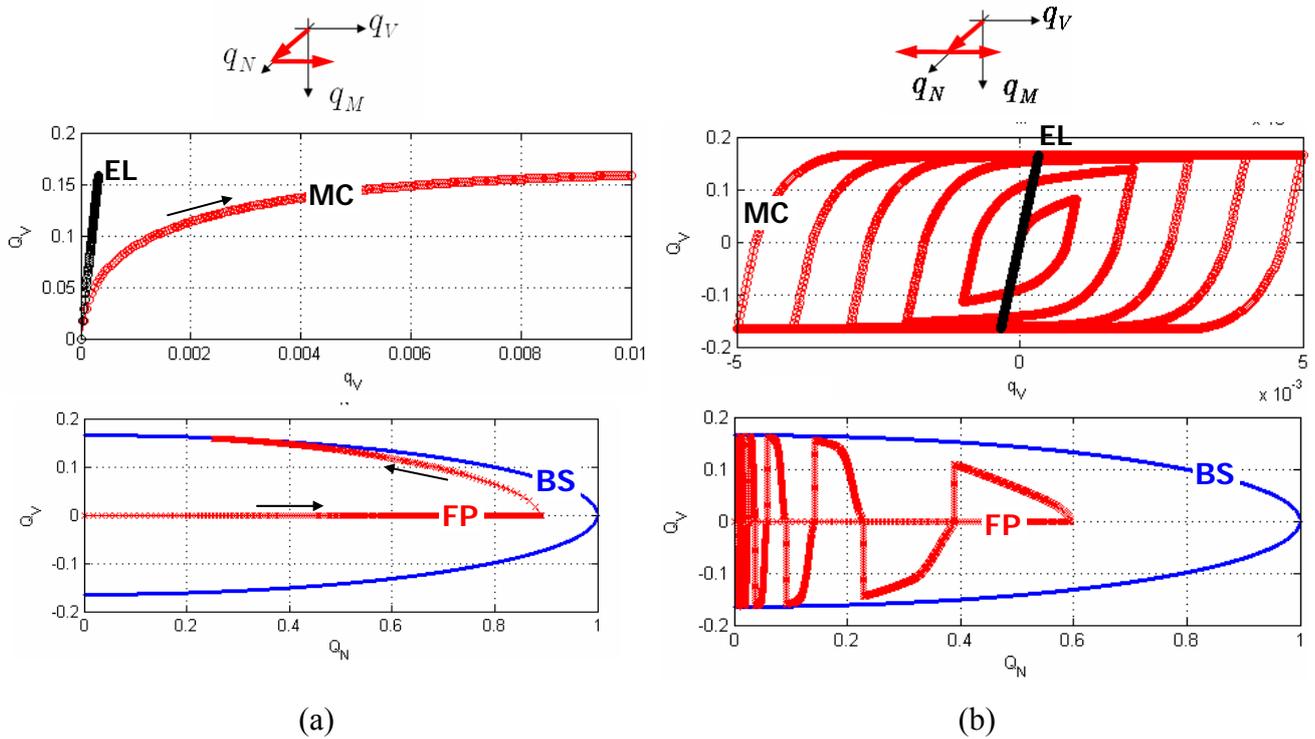

(a)  (b)

**EL**: *elastic response,* **MC**: *full macroelement response,* **BS**: *bounding surface,* **FP**: *recorded force path.*

*Figure 9 – System response under quasistatic horizontal loading: (a) Monotonic loading, (b) Cyclic loading with increasing loading amplitude.*

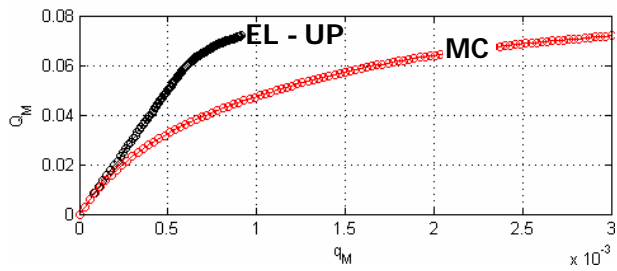
(a-1)
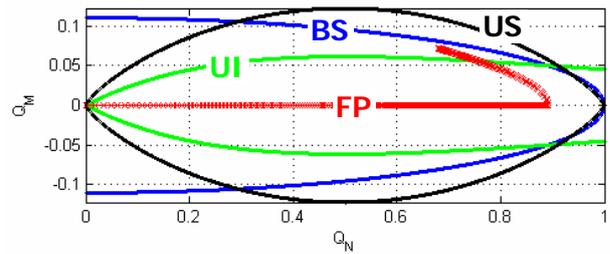
(a-2)

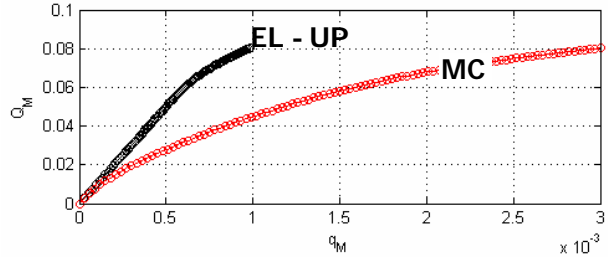
(b-1)
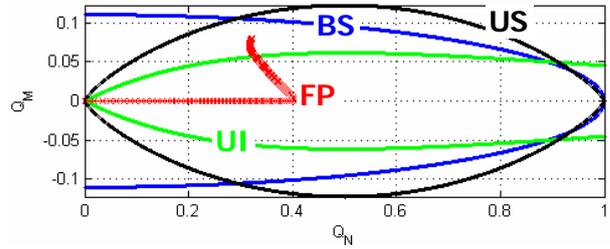
(b-2)

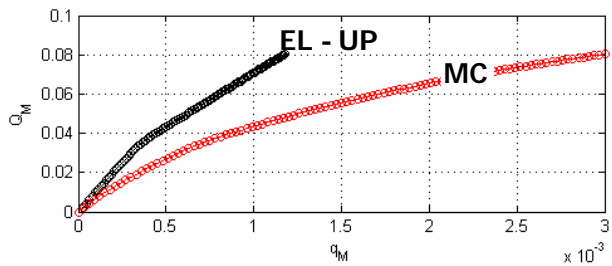
(c-1)
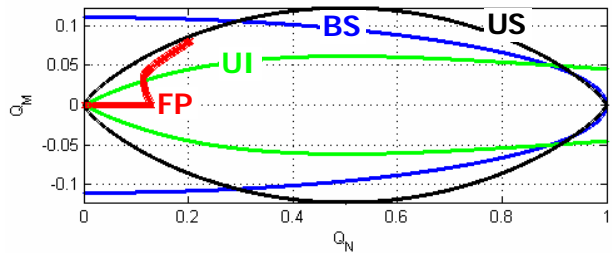
(c-2)

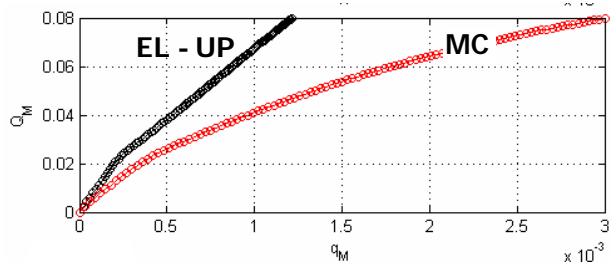
(d-1)
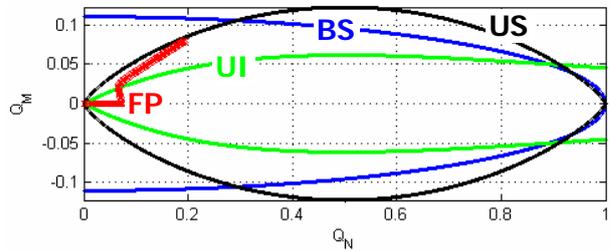
(d-2)

**EL-UP**: *non linear elastic response* (*with uplift*), **MC**: *full macroelement response*, **BS**: *bounding surface*, **FP**: *recorded force path,* **US**: *ultimate surface,* **UI**: *surface of uplift initiation*

*Figure 10 – System response under quasistatic monotonic moment loading*

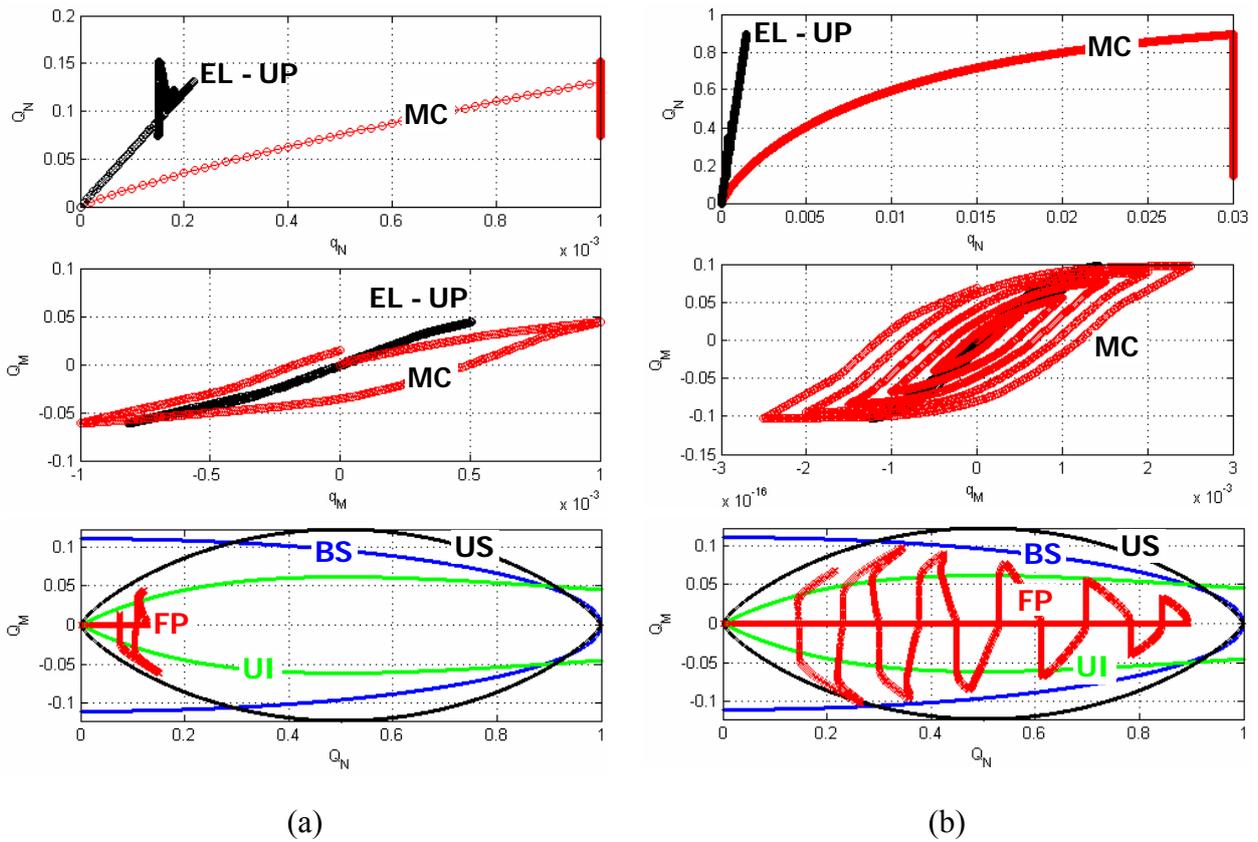

(a)                          (b)

**EL-UP**: *non linear elastic response* (*with uplift*), **MC**: *full macroelement response*, **BS**: *bounding surface*, **FP**: *recorded force path*, **US**: *ultimate surface*, **UI**: *surface of uplift initiation*

*Figure 11 – System response under quasistatic cyclic moment loading: (a) One single cycle with uplift and (b) Five successive increasing cycles.*

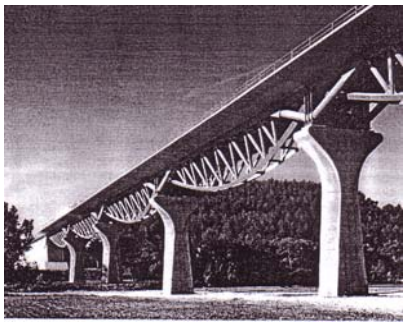
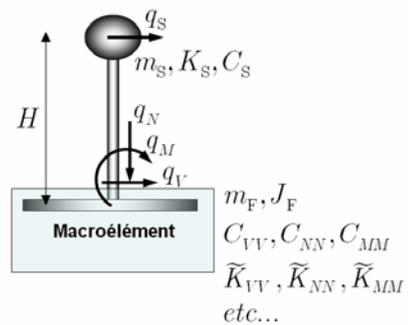

(a)                          (b)

*Figure 12 –* (a) *The columns of the Arc Viaduct in France (Design by Greisch Consultants) and* (b) *Simple model for dynamic analysis.*

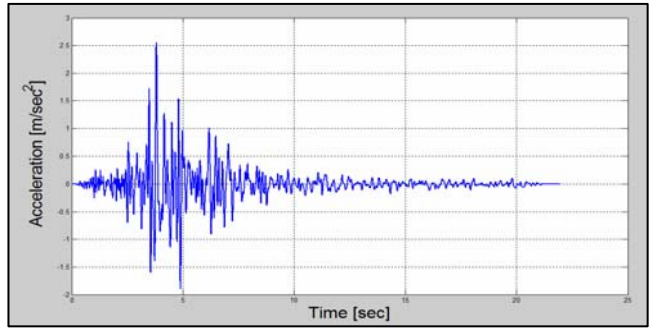

*Figure 13 – Acceleration time history recorded during the Friuli earthquake (Italy, 1976)*

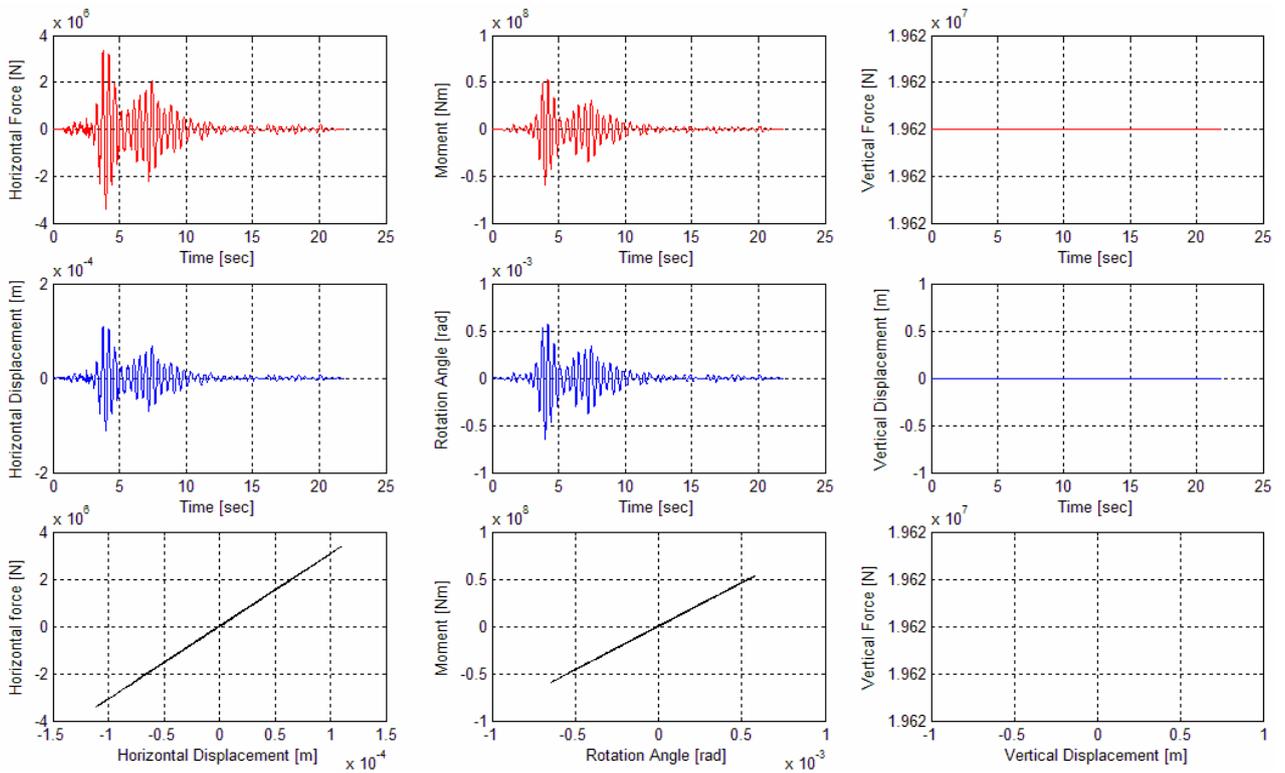

*Figure 14 – Response of the system considering linear elastic behaviour*

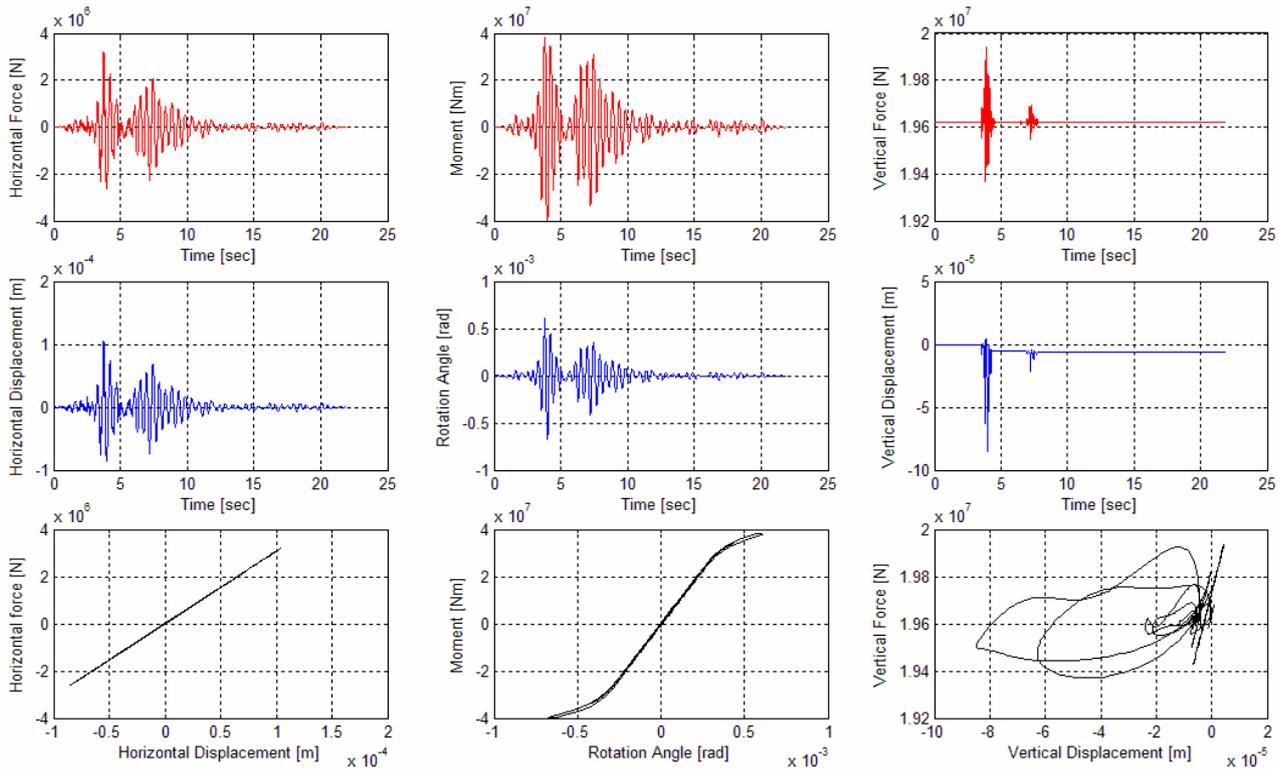

*Figure 15 – System response considering purely elastic behaviour with uplift*

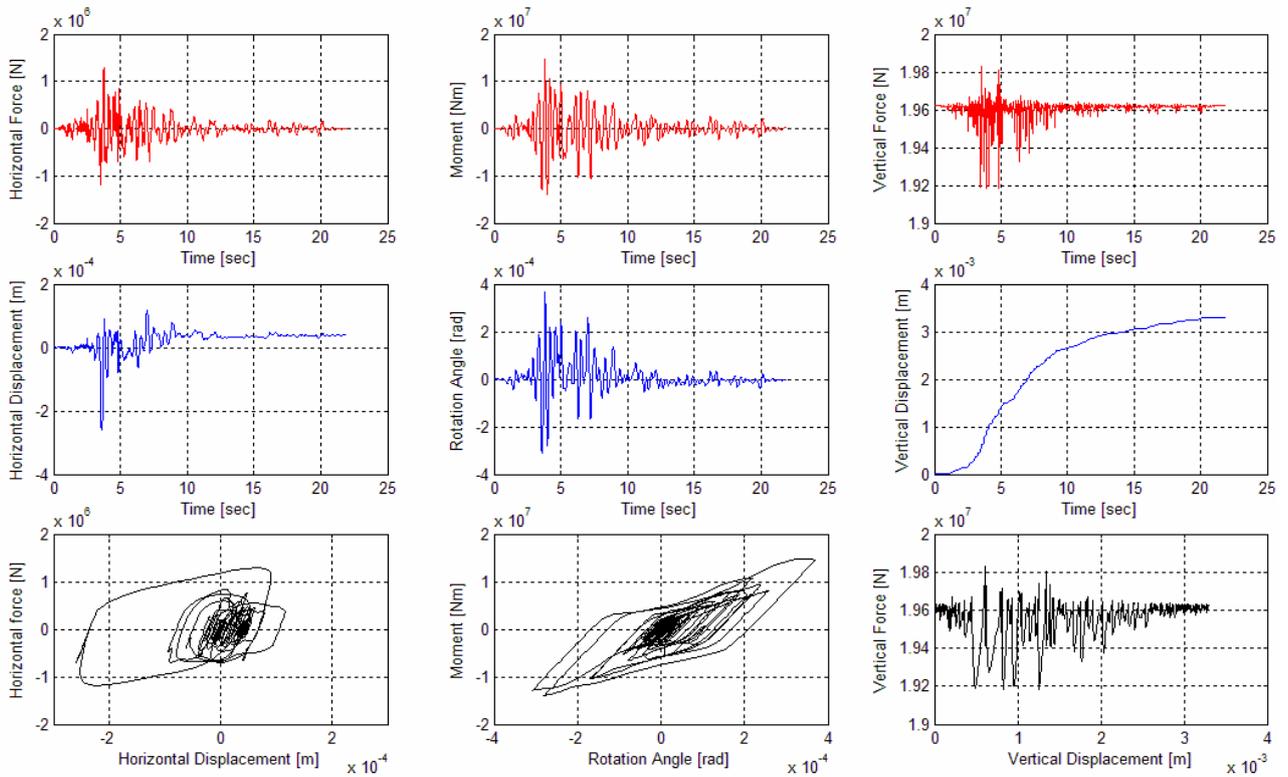

*Figure 16 – System response considering fully elastoplastic behaviour with uplift*

# List of Tables

*Table 1 – Overview of existing macroelement models for shallow foundations*

| Reference | Year | Configuration | Description |
|---|---|---|---|
| **Nova & Montrasio** | 1991 | Strip footing resting on a purely frictional soil | Isotropic hardening plasticity model and non-associated flow rule. Application in the case of quasistatic monotonic loading. |
| **Paolucci** | 1997 | Strip footing resting on a purely frictional soil | Perfect plasticity model with non-associated flow rule. Application to simple structures subject to seismic loading. Parametric studies. |
| **Pedretti** | 1998 | Strip footing resting on a purely frictional soil | Hypoplastic model for the description of the system response under cyclic loading. Consideration of uplift by reduction of the elastic stiffness. Applications to structures subject to quasistatic cyclic loading. |
| **Gottardi *et al.* (*cf.* [12])** | 1999 | Strip footing resting on a purely frictional soil | Isotropic hardening plasticity model. Detailed description of the system ultimate surface (identified as the yield surface of the plasticity model) via "swipe tests". Application in the case of quasistatic monotonic loading. |
| **Le Pape *et al.* Le Pape & Sieffert** | 1999 2001 | Strip footing resting on a purely frictional soil | Elastoplastic model derived from thermodynamical principles. Rugby ball shaped yield surface and ellipsoidal plastic potential. Application to seismic loading. |
| **Crémer *et al.*** | 2001, 2002 | Strip footing resting on a purely cohesive soil without resistance to tension | Non-associated plasticity model with isotropic and cinematic hardening coupled with a model for uplift. Application to seismic loading. |
| **Martin & Houlsby (*cf.* [13])** | 2001 | Circular footing resting on a purely cohesive soil | Non-associated plasticity model with isotropic hardening. Detailed description of the yield surface via "swipe tests". Application to quasistatic monotonic loading. |
| **Houlsby & Calssidy (*cf.* [14])** | 2002 | Circular footing resting on a purely frictional soil | Non-associated plasticity model with isotropic hardening. Detailed description of the yield surface via "swipe tests". Application to quasistatic monotonic loading. |
| **Di Prisco *et al.* (*cf.* [15])** | 2003 | Strip footing resting on a purely frictional soil | Hypoplastic model for the description of the behavior under cyclic loading. Application to quasistatic cyclic loading. |
| **Cassidy *et al.* (*cf.* [16])** | 2004 | Circular footing resting on a frictional or cohesive soil | Fully three-dimensional formulation. Application to the off-shore industry. Quasistatic monotonic loading. |
| **Houlsby *et al.*** | 2005 | Strip or circular footing resting on cohesive soil. Frictional soil-footing interface | Decoupled Winkler springs with elastic perfectly plastic contact-breaking law derived from thermodynamical principles. Application to quasistatic cyclic loading. |
| **Einav & Cassidy** | 2005 | Strip footing resting on cohesive soil. Frictional soil-footing interface | Decoupled Winkler springs with elastoplastic contact-breaking law with hardening derived from thermodynamical principles. Application to quasistatic cyclic loading. |
| **Grange *et al.* (*cf.* [17])** | 2006 | Circular footing on cohesive soil | Extension of the plasticity model of Crémer to purely three-dimensional setting. No separate uplift model included. |

*Table 2– Approximate relationships for the Static impedances and the radiation damping coefficients of a circular footing resting on homogeneous elastic half-space*

| Mode | Static Impedances | Radiation damping coefficients |
|---|---|---|
| Vertical | $\widetilde{K}_{NN} = \dfrac{4\mathcal{G}r}{1-\nu}$ | $C_{NN} = \rho V_{\text{La}}\left(\pi r^2\right)$ |
| Horizontal | $\widetilde{K}_{VV} = \dfrac{8\mathcal{G}r}{2-\nu}$ | $C_{VV} = \rho V_{\text{S}}\left(\pi r^2\right)$ |
| Rotation | $\widetilde{K}_{MM} = \dfrac{8\mathcal{G}r^3}{3(1-\nu)}$ | $C_{MM} = \rho V_{\text{La}}\left(\dfrac{\pi r^4}{4}\right)$ |